\documentstyle[12pt,amsmath,cite]{article}

\evensidemargin=\oddsidemargin
\newcommand{\be}{\begin{equation}}
\newcommand{\ee}{\end{equation}}
\newcommand{\bea}{\begin{eqnarray}}
\newcommand{\eea}{\end{eqnarray}}

\newcommand{\p}{\partial}

\def\half{\frac{1}{2}}

\def\IB{\relax\hbox{$\inbar\kern-.3em{\rm B}$}}
\def\IC{\relax\hbox{$\inbar\kern-.3em{\rm C}$}}
\def\ID{\relax\hbox{$\inbar\kern-.3em{\rm D}$}}
\def\IE{\relax\hbox{$\inbar\kern-.3em{\rm E}$}}
\def\IF{\relax\hbox{$\inbar\kern-.3em{\rm F}$}}
\def\IG{\relax\hbox{$\inbar\kern-.3em{\rm G}$}}
\def\IGa{\relax\hbox{${\rm I}\kern-.18em\Gamma$}}
\def\IH{\relax{\rm I\kern-.18em H}}
\def\IK{\relax{\rm I\kern-.18em K}}
\def\IL{\relax{\rm I\kern-.18em L}}
\def\IP{\relax{\rm I\kern-.18em P}}
\def\IR{\relax{\rm I\kern-.18em R}}
\def\IZ{\relax{\rm Z\kern-.5em Z}}

\def\tr{\hbox{tr}}

\begin{document}

\begin{titlepage}

\begin{flushright}
GEF-TH-08/2006\\
SNUTP/06-005\\
hep-th/0605278\\
\end{flushright}

\vskip .1in

\begin{center}

{\Large\bf Time-dependent supergravity solutions in null
dilaton background}

\vskip .6in
{\bf {Rashmi R. Nayak$^a$ \footnote{email: Rashmi.Nayak@ge.infn.it},
Kamal L. Panigrahi$^{a,b}$ \footnote{email: Kamal.Panigrahi@ge.infn.it,
panigrahi@iitg.ernet.in} and Sanjay Siwach$^c$ \footnote{e-mail:
{sksiwach@phya.snu.ac.kr}}}}

\vskip .1in
{$^a$\it Dipartimento di Fisica,
Universita' \& INFN Sezione di Genova, ``Genova'' \\
Via Dodecaneso 33, 16146, Genova, Italy}\\
\vskip .1in
{$^b$\it Physics Department,
Indian Institute of Technology, Guwahati, 781 039, India}\\
\vskip .1in
{$^c$\it Centre for Theoretical Physics, Seoul National University,
Seoul, 151-747, Korea}
\vskip .2in

\vskip .4in

\begin{abstract}
\vskip .1in
A class of time dependent pp-waves with NS-NS flux in type IIA string theory
is considered. The background preserves 1/4 supersymmetry and may provide a
toy model of Big Bang cosmology with non trivial flux. At the Big Bang
singularity in early past, the string theory is strongly coupled and Matrix
string model can be used to describe the dynamics. We also construct some time
dependent supergravity solutions for D-branes and analyze their supersymmetry
properties.

\end{abstract}

\end{center}

\vfill

\end{titlepage}
\section{Introduction}

\noindent
The study of time dependent backgrounds in string theory is a challenging
problem. Until now quite a few time dependent solutions in string
theory/supergravity are known. Recently a simple cosmological background with
a null dilaton has been proposed \cite{Craps:2005wd} as a toy model of Big
Bang cosmology. The background preserves half of the thirty-two supersymmetries
and one expects that there is a good control over the singularity at early
times. The matrix degrees of freedom, rather than point particles or the
perturbative string states describe the correct physics near the singularity.
In discrete light cone type of quantization, a matrix string
\cite{Banks:1996my,Dijkgraaf:1997vv,Bonelli:2002mb} with a time dependent
coupling
constant is used to study the dynamics at the Big Bang. The success of the
model has to be tested by the further investigation of the relevant matrix
model and its cosmological predictions. Some steps have already been taken
in this direction
\cite{Li:2005sz,Berkooz:2005ym,Li:2005ti,Hikida:2005ec,Li:2005ai,Tai:2006my,
Das:2005vd,
Kawai:2005jx,She:2005mt,Chen:2005bk,Chen:2005mg,Robbins:2005ua,
KalyanaRama:2005uw,Ishino:2005ru,
Ishino:2006nx,Chu:2006pa,Das:2006dr,Das:2006dz,Lin:2006ie,Martinec:2006ak,
Craps:2006xq,Chen:2006rm,Kodama:2006bw}
and the future studies may seed the revolution for the string cosmology. See
\cite{Craps:2006yb} for a recent review of the subject.

A non-trivial extension of the background is to include the RR and NS-NS
fluxes and some time dependent pp-wave solutions of the string theory with a null
dilaton are considered in the literature \cite{Chen:2005bk} and the matrix string theory of a class of
pp-wave background has been analyzed \cite{Das:2005vd,Das:2006dr} (see also
\cite{Papadopoulos:2002bg,Blau:2003rt} for some early
studies of time dependent pp-waves).
The AdS/CFT correspondence has also been
considered for a time dependent type IIB background\cite{Chu:2006pa,Das:2006dz,
Lin:2006ie,Martinec:2006ak} and the dual gauge theory has been realized as a time
dependent supersymmetric Yang-Mills theory living on the boundary. In AdS/CFT
correspondence, supersymmetry plays a very important role and the existence of
supersymmetry gives us a better control over the nonperturbative behavior of
the dual quantum field theory. In this sense, the time dependent backgrounds
preserving some supersymmetry are useful tools for
the study of the quantum gravity and dual field theories. The study of open
strings in time dependent backgrounds is also an interesting problem. In this
context, some D-brane solutions in a light-like linear dilaton background has
also been discussed in \cite{Nayak:2006dm} by taking a suitable Penrose limit
of an intersecting brane solution in supergravity.

In this paper, we study the type IIA pp-waves with NS-NS three form flux in a
light-like linear dilaton background. This can be considered as the simplest
extension of the model of \cite{Craps:2005wd} with non-trivial flux.
We study the supersymmetry of the background and show that it preserves 1/4 of type IIA supersymmetry.
At early times, the string theory is strongly coupled and the matrix string description may be used.
We also construct some time dependent supergravity p-brane solutions preserving some
fraction of supersymmetry in this background. The solutions have singularities
at early times similar to the Big Bang model but perhaps the matrix degrees of freedom
may be used to resolve the singularity.

The rest of the paper is organized as follows. In section-2 we study the
pp-waves with three form NS-NS flux as a toy model of Big Bang cosmology.
We analyze the geodesic equations in this background and also give the matrix
string description of this class of solutions.
We find that the background preserves 1/4 of the type II supersymmetry.
In section-3, we present classical solutions of D-branes and analyze
the supersymmetry properties of these solutions. We show that they preserve 1/8
supersymmetry. Finally in section-4, we present our conclusions.

\section{The type II pp-waves with null dilaton}

We begin with the following ansatz for the background metric and
other fields,
\bea
ds^2 &=& -2 dx^{+}dx^{-} - \mu^2 (x^{+}) \sum_i (x^i)^2 dx^{+}dx^{+}
+ dx^i dx^i + dy^a dy^a, \cr
&\cr
 \Phi &=& - Q x^{+}, \>\>\> \>\>\> H_{+12} = H_{+34} = 2f (x^{+}).
\label{sugra} \eea Note that the dilaton is linear in light cone
time and the indices run as $ i = 1,..., 4,\>\>\> a = 5,...,8 $.
Only non zero component of Ricci curvature computed from the above
metric is, \be R_{++} = 4 \mu^2 (x^{+}) \ee and to be a solution
of type IIA supergravity equations of motion, one should satisfy
\be \mu^2 = -\half \ddot\Phi +  f^2 = f^2, \ee where dots denotes
derivative with respect to $x^+$. At the Big Bang, $x^+
\rightarrow -\infty$  the dilaton diverges and string theory is
strongly coupled and a DLCQ type of description can be given (see
below). To study the nature of singularity at early times let us
consider the geodesic motion of a point particle in our
background. For this we change our metric to the Einstein frame
metric, \bea ds^2_{\rm E} = e^{Qx^+/2}\left[-2 dx^{+}dx^{-} -
\mu^2 (x^{+}) (x^i)^2 dx^{+}dx^{+} + dx^i dx^i + dy^a dy^a\right].
\eea At $x^+ \rightarrow -\infty$, the metric components shrink to
zero, which corresponds to Big Bang singularity. We would like to
study the geodesic equation for a test particle near the
singularity. Null geodesic in the spacetime at constant $X^-, x^i$
and at $X^a = 0$ is given by \bea \frac{d^2 X^+}{d\sigma^2} +
\Gamma^{+}_{\alpha \beta}
\frac{dx^{\alpha}}{d\sigma}\frac{dx^{\beta}}{d\sigma} = 0. \eea
which, for our metric, can be written as \bea \frac{d^2
X^+}{d\sigma^2} + \frac{Q}{2}
\left(\frac{dx^{+}}{d\sigma}\right)^2 = 0. \eea Solving the above
equation, one gets the affine parameter \bea \sigma = {\rm exp}(Q
x^+/2), \eea upto a reparameterization. Therefore the singularity
at $x^+ \rightarrow -\infty$ correspond to $\sigma =0$, and it has
finite affine distance to all points interior of the spacetime and
the spacetime is geodesically incomplete. One can try to compute
the Riemann tensors and can show that there is a curvature
singularity at $\sigma =0$ and gives a divergent tidal force felt
by the inertial observer. At  late times, $x^+ \rightarrow \infty$
the affine parameter diverges and it corresponds to the asymptotic
region of spacetime.

\vskip .5cm
\noindent
{\bf {Supersymmetry:}} The supersymmetry variation of dilatino and
gravitino in string frame is given by
\bea
\delta \lambda &=& \half(\Gamma^{\mu}\partial_{\mu}\phi -
\frac{1}{12} \Gamma^{\mu \nu \rho}H_{\mu \nu \rho}\Gamma_{11})
\epsilon + \cdots \\
\label{dilatino}
\delta \Psi_\mu &=& \Big[\partial_{\mu} + \frac{1}{4}(w_{\mu
  \hat a \hat b} - \half H_{\mu \hat{a}
  \hat{b}}\Gamma_{11})\Gamma^{\hat{a}\hat{b}}\Big]\epsilon + \cdots
\label{gravitino}
\eea
where the dots stand for the terms coming for the R-R charges,
and we have used $(\mu, \nu ,\rho)$ to
describe the ten dimensional space-time indices, and hated indices represent
the flat tangent space indices.

Veilbeins and spin connection for the above metric are,
\bea
e^{\hat +}_+ &=& 1, \>\>\> e^{\hat -}_- = 1,\>\>\>e^{\hat -}_+
= \half \mu^2 x^2_i   ,\>\>\> e^{\hat i}_j = \delta^{\hat i}_j \cr
&\cr
&&e^{\hat a}_b = \delta^{\hat a}_b, \>\>\> \omega^{~\hat -\hat i}_+
= \mu^2 x^i
\eea
For our metric and linear null dilaton background, putting the
supersymmetry variations of dilatino and gravitino equal zero to gives,
\bea
\delta \lambda &\equiv&\left(-Q \Gamma^{\hat +} - \frac{1}{12}
\Gamma^{\hat + \hat i \hat j}
H_{+ij}\Gamma_{11}\right)\epsilon = 0, \cr
&\cr
\delta \Psi_+ &\equiv& \left(\p_+ - \half \mu^2 x^i \Gamma^{\hat + \hat i}
- \frac{1}{2} \mu (x^+) (\Gamma^{\hat 1 \hat 2} +\Gamma^{\hat 3 \hat 4})
\Gamma_{11} \right)\epsilon = 0, \cr
&\cr
\delta \Psi_- &\equiv& \p_- \epsilon = 0, ~~~~~\delta \Psi_a = \p_a
\epsilon = 0,\cr
&\cr
\delta \Psi_i &\equiv& \left(\p_i - \frac{1}{8} H_{+ij}\Gamma^{\hat + \hat j}
\Gamma_{11}\right)\epsilon = 0
\eea
The dilatino variation is solved by imposing
\be
\Gamma^{\hat +} \epsilon = 0,
\ee
which breaks half of the supersymmetry. The $\delta \Psi_+ = 0$ equation
further requires the condition
\be
(\Gamma^{\hat 1 \hat 2} +\Gamma^{\hat 3 \hat 4}) \epsilon = 0
\ee
which breaks another half of the supersymmetry. Thus the background preserve
1/4 supersymmetry.

\vskip .5cm
\noindent
{\bf{Matrix string description:}}
Near the big bang singularity, the dilaton diverges and the string theory is
strongly coupled. However one can give a discrete light cone type of quanntization (DLCQ) at the early
times. For the DLCQ of the background, we pick up the direction $y^8$ and make
the following identifications,
\be
(x^+, x^-, y^8) \sim (x^+, x^- + R, y^8 + \epsilon R).
\ee
Making a Lorentz transformation as in \cite{Craps:2005wd}
\bea
x^+ &=& \epsilon X^+ , ~~~~
x^- = \frac{X^+}{2\epsilon} + \frac{X^-}{\epsilon} + \frac{Y^5}{\epsilon} \cr
&\cr
y^8 &=& X^+ + Y^8, ~~~~~
x^{1,2,3,4} = X^{1,2,3,4}, \>\>\>\> y^{5,6,7} = Y^{5,6,7}
\eea
and applying the T duality along $Y^8$ followed by an S duality we get the
IIB background,
\bea
ds^2 &=& r e^{\epsilon Q X^+}\left(-2 dX^{+}dX^{-} - \mu^2 (X^{+}) \epsilon^2
(X^i)^2 dX^{+}dX^{+} + dX^idX^i + dY^adY^a \right) \cr
&\cr
\Phi &=&  Q \epsilon X^{+} + \log r, \>\>\> F_{+12} = F_{+34} = 2 f (X^{+}),
\>\>\> i = 1,...,4, \>\>\> a = 5,...,8
\eea
where $r = \frac{\epsilon R}{2 \pi l_s}$ and $Y^8 \sim Y^8 + \frac{2 \pi l_s}
{r}$.

Let us focus on the DBI action of a single D1-brane in this
background, \be S_{D1} = - \frac{1}{2 \pi l^2_s} \int d\tau
d\sigma e^{-\Phi} \sqrt{-det\left (\p_\alpha X^\mu\p_\beta X^\nu
G_{\mu\nu} + 2 \pi l^2_s F_{\alpha\beta}\right)} \ee Let us take
$F_{\alpha\beta}=0$, $Y^8 = \frac{\sigma}{r}$ and  $X^+$ and $X^-$
depending on $\tau$ only. The equations of motion derived from the
above action require, $\p_\tau X^+ =\p_\tau X^-$ and a classical
solution is given by, \be X^+ =
\frac{1}{r}\frac{\tau}{\sqrt{2}},~~~X^- = \frac{1}{r}\frac{\tau}
{\sqrt{2}},~~~~Y^8 = \frac{1}{r} \sigma,~~~~Y^{5,6,7} = 0, \>\>\>
X^i = 0. \ee Choosing the gauge, $X^+ =
\frac{1}{r}\frac{\tau}{\sqrt{2}},~~~ Y^8 = \frac{\sigma}{r}$ and
defining the new variable $Z$ as, $X^- =
\frac{1}{r}\frac{\tau}{\sqrt{2}} + \sqrt{2} Z$, and expanding the
action around the classical solution upto quadratic terms in the
fluctuations, we get, \bea S_{D1} &=&  \frac{1}{2 \pi l^2_s} \int
d\tau d\sigma \Big(-\frac{1}{r^2} + \half \frac{(\mu
\epsilon)^2}{r^2}(X^i)^2 + \half[\p_\tau Z)^2 - (\p_\sigma Z)^2 +
(\p_\tau X^i)^2 - (\p_\sigma X^i)^2 \cr & \cr &+& (\p_\tau
Y^{5,6,7})^2 - (\p_\sigma Y^{5,6,7})^2 ] + 2\pi^2 l^4_s
\exp{(-\frac{\sqrt{2}\epsilon Q\tau}{r})} F^2_{\tau\sigma}\Big).
\eea This is the action for a single D1 branes that follows from
the matrix string action \cite{Banks:1996my,Dijkgraaf:1997vv} for
N D1 branes, \bea S_{D1} &=&  \frac{1}{2 \pi l^2_s} \int \tr
\Big(\half (D_\mu X^i)^2 + \half (D_\mu Y^a)^2 + \half \frac{(\mu
\epsilon)^2}{r^2}(X^i)^2 + g^2_s l^4_s \pi^2 F^2_{\mu\nu} \cr &
\cr &-& \frac{1}{4\pi^2 g^2_s l^4_s} [X^i, X^j]^2 -
\frac{1}{4\pi^2 g^2_s l^4_s} [Y^a, Y^b]^2 - \frac{1}{4\pi^2 g^2_s
l^4_s} [X^i, Y^a]^2  + ...\Big), \eea where the dots stand for
fermionic terms.

\section{D-branes solutions}

In general it is a difficult task to find p-brane solutions of supergravity
in time dependent backgrounds. However in the linear dilaton background the
following factorized ansatz solve the type II supergravity equations of motion.
\bea
ds^2 &=& e^{\frac{f}{2}} H^{-\half}
\left[-2 dx^+ dx^- - \mu^2 (x^+) x^2_i (dx^+)^2 +
d\vec y^2_\alpha \right] + e^{-\frac{f}{2}}
H^\half {d\vec x_a}^2 \cr
& \cr
e^{2\Phi} &=& e^{-\frac{(7-p)}{2}f} H^{\frac{(3-p)}{2}},
\>\>\>\>H_{+12} = H_{+34} = 2\mu (x^+)\cr
&\cr
F_{+ - \alpha...a} &=&  e^{2f} \p_a H^{-1},
~~~~({\rm for}~ p\le 3)\cr
&\cr
F_{a_1...a_n} &=& \epsilon_{a_1...a_n a} \p_a H,  ~~~~ ({\rm for}~ p\ge 3)
\eea
where $i =x^1,\cdots,x^4,{\rm and}\>\>$, and $x^{\alpha}$ and  $x^{a}$
correspond
to the directions parallel and transverse to the Dp-brane respectively.
$f = Q x^{+}$, is a function of $x^+$ only and
$ H = 1 + (\frac{r_0}{r})^{\tilde d}$, (where $\tilde d = 7-p$
in ten dimensions), is the harmonic function in transverse
space and is independent of $x^+$.

However, asymptotically the above solutions
do not go to the pp-wave background with null linear dilaton discussed in the
previous section. To remedy the
situation one has to give up the factorized ansatz and replace the harmonic
function in the transverse space by
$H \rightarrow 1 + e^{-Q x^+}(\frac{r_0}{r})^{\tilde d}$ and modify and
rewrite the field strengths accordingly. Explicitly one can check that the
following solutions obey the supergravity equations of motion,
\bea
ds^2 &=& \tilde H^{-\half}\left[-2 dx^+ dx^- - \mu^2 (x^+) x^2_i (dx^+)^2 +
d\vec y^2_\alpha \right] + \tilde H^\half {d\vec x_a}^2 \cr
& \cr
e^{2\Phi} &=& e^{- 2 f} \tilde H^{\frac{(3-p)}{2}}, \>\>\>\>H_{+12}
= H_{+34} = 2\mu (x^+)\cr
&\cr
F_{+-\alpha...a} &=&  e^{f} \p_a \tilde H^{-1},
~~~~({\rm for}~ p\le 3)\cr
&\cr
F_{a_1...a_n} &=& \epsilon_{a_1...a_n a}e^f \p_a \tilde H,
~~~~ ({\rm for}~ p\ge 3)
\label{Dp-brane}
\eea
where $f = Q x^{+}$ as before and
$\tilde H = 1 + e^{-Qx^+}(\frac{r_0}{r})^{\tilde d}$.

We would like to study the geodesic equations of a point particle in
this background. To see the nature of the trajectory, we pass on to
the Einstein frame and the metric for the D$p$-brane is given
by,
\bea
ds^2_{\rm E} = e^{f/2}\tilde H^{-\frac{7-p}{8}}
\left[-2 dx^+ dx^- - \mu^2 (x^+) x^2_i (dx^+)^2 +
d\vec y^2_\alpha \right] + e^{f/2} \tilde H^{\frac{p+1}{8}}
{d\vec x_a}^2 \ .
\eea
We would like to examine the trajectories near the singularity at constant
$X^-, y^{\alpha}$ and at $X^{a} = 0$. Namely we check the
trajectories moving along the $X^+$, which is given by
\bea
\frac{d^2 X^+}{d\sigma^2} + \left(\frac{\dot f}{2} +
\frac{p-7}{8}\frac{\dot{\tilde H}}{\tilde H}\right)
\frac{dx^{+}}{d\sigma}\frac{dx^{+}}{d\sigma} = 0.
\eea
Near the Big bang singularity (at $x^+ \rightarrow -\infty$), the above
equation can be written as,
\bea
\frac{d^2 X^+}{d\sigma^2} + c
\frac{dx^{+}}{d\sigma}\frac{dx^{+}}{d\sigma} = 0.
\eea
where, $f = Qx^+$ and $c= ((11-p)/8)Q$. Hence the affine parameter is
\bea
\sigma = e^{cx^+}
\eea
upto an affine reparameterization invariance. The singularity
at $x^+ \rightarrow -\infty$ correspond to $\sigma =0$, and can be approached
in a finite affine time and the spacetime is geodesically incomplete.
On the other hand, at $x^+ \rightarrow \infty$, the string coupling $g_s$
goes to zero, and the spacetime theory is free at late times. So the
geodesic behavior is qualitatively similar to that of our background in the
previous section.
\vskip .5cm
\noindent
{\bf {Supersymmetry:}}
 Next we would like to analyze the supersymmetry of the Dp-brane
solutions presented above. The supersymmetry variation of the
gravitino and dilatino fields in type IIA supergravity in string frame is given by
\cite{Schwarz:1983qr,Hassan:1999bv},
\begin{eqnarray}
\delta \lambda = \half(\Gamma^{\lambda}\partial_{\lambda}\Phi -
\frac{1}{12} \Gamma^{\mu \nu \rho}H_{\mu \nu \rho} \Gamma_{11})
\epsilon + \frac{1}{8}e^{\Phi}(5 F^{0} - \frac{3}{2!} \Gamma^{\mu \nu}
  F^{(2)}_{\mu \nu} \Gamma_{11} + \frac{1}{4!} \Gamma^{\mu \nu \rho \sigma}
  F^{(4)}_{\mu \nu \rho \sigma})\epsilon, \nonumber \\
\label{dilatino-IIA}
\end{eqnarray}
\begin{eqnarray}
\delta {\Psi_{\mu}} = \Big[\partial_{\mu} + \frac{1}{8}\omega_{\mu
 \hat a \hat b} \Gamma^{\hat a \hat b}- \frac{1}{8} H_{\mu \hat{a}
 \hat{b}}\Gamma^{\hat{a}\hat{b}}\Gamma_{11}\Big]\epsilon
+ \frac{e^{\Phi}}{8}\Big[ F^{(0)} - \frac{1}{2!}
\Gamma^{\mu \nu} F^{(2)}_{\mu \nu} \Gamma_{11}+ \frac{1}{4!}
\Gamma^{\mu \nu \rho \sigma}F^{(4)}_{\mu \nu \rho \sigma}\Big]\Gamma_{\mu}
\epsilon , \nonumber \\
\label{gravitino-IIA}
\end{eqnarray}
where we have used $(\mu, \nu ,\rho)$ to describe the ten dimensional
space-time indices, and the hated indices are the corresponding tangent
space indices. Solving the dilatino variations for the Dp-brane
solutions presented in (\ref{Dp-brane}) gives the following
conditions on the spinor,
\bea
\Gamma^{\hat +}\epsilon =0,
\label{Dpgammaplus}
\eea
and
\bea
&&\Gamma^{\hat a}\epsilon - \Gamma^{\hat + \hat - \hat \alpha_2 \cdots
\hat \alpha_p \hat a}
\epsilon =0 \>\>\> ({\rm for} \>\> p \le 3) \cr & \cr
&&\Gamma^{\hat a}\epsilon + \frac{1}{(n+1)!}\epsilon_{\hat a_1 \cdots
\hat a_n \hat a}
\Gamma^{\hat a_1 \hat a_2 \cdots \hat a_n \hat a} \epsilon =0 \>\>\>\>
({\rm for} \>\>\> p \ge 3) .
\label{Dpbranesusy}
\eea
We would like to emphasize that we need to impose both the conditions,
(\ref{Dpgammaplus}) and (\ref{Dpbranesusy}), for the
dilatino variation to be satisfied. Next we would like analyze the
gravitino variation,
\bea
\delta \Psi_+ &\equiv& \Big(\p_+ - \frac{1}{8}\frac{\p_+ \tilde H}
{\tilde H}  - \frac{\mu}{2}
\tilde H^{-\half}(\Gamma^{\hat 1 \hat 2}+ \Gamma^{\hat 3 \hat 4})\Gamma_{11}
\Big)\epsilon = 0 ,\cr
&\cr
\delta \Psi_- &\equiv& \p_- \epsilon = 0 , ~~~~
\delta \Psi_\alpha \equiv \p_\alpha \epsilon = 0 ,\cr
&\cr
\delta \Psi_a &\equiv&\Big(\p_a  - \frac{1}{8}\frac{\p_a \tilde H}
{\tilde H}  \Big)\epsilon = 0 ,
\eea
where in writing the above variations we have used the condition
(\ref{Dpgammaplus}) and the brane supersymmetry condition (\ref{Dpbranesusy}).
Now using the condition,
\be
(\Gamma^{\hat 1 \hat 2} + \Gamma^{\hat 3 \hat 4}) \epsilon = 0 ,
\label{Dpgamma}
\ee
we are left with the following equations to be solved
\bea
\left(\p_+ - \frac{1}{8}\frac{\p_+ \tilde H}{\tilde H}\right)\epsilon = 0,
\>\>\>
\left(\p_a  - \frac{1}{8}\frac{\p_a \tilde H}{\tilde H}\right)\epsilon = 0.
\eea
The above equations are solved by the spinor of the form
$\epsilon = \tilde H^{\frac{1}{8}}\epsilon_0$, with $\epsilon_0$ being
a constant spinor. Now putting together the conditions,
(\ref{Dpgammaplus}) (\ref{Dpbranesusy}), and (\ref{Dpgamma}), we conclude
that our solutions (\ref{Dp-brane}) preserves 1/8 supersymmetry.

\section{Conclusions}
In this paper, we discussed a class of time dependent pp-waves
with NS-NS three form flux in  null linear dilaton background. It
may serve as a toy model of Big Bang cosmology with non-trivial
flux. At the beginning of the time, the dilaton diverges and the
singularity may be interpreted as Big Bang singularity at the
origin of the time. The background preseve 1/4 supersymmetry as
opposed to 1/2 supersymmetry of the background in the absence of
the flux. The geodesic equation near the singularity has been
analyzed. Near the singularity, the perturbative string theory
fails as dilaton is divergent, however one can use a discrete
light cone type of quantization and the corresponding matrix
string description is given.

We have also constructed a class of D-brane solutions in this background, by
solving the type II field equations explicitly. Though we restrict ourselves
to the type IIA brane solutions in this paper, the general exprssion is also
valid for p-branes in  type IIB theory as well. We have also worked out the
geodesic equations, and the nature of singularity at early times.
The supersymmetry variation revealed that this class of branes preseve 1/8
of the full typs IIA spacetime supersymmetry.

There are several directions for future work. The lift of the
background to eleven dimensional M-theory is straightforward. One
can extend the discussion with RR three form flux and
corresponding S-dual type IIB description can be written down. The
list of D-brane solutions given by us, is based on a clever ansatz
and perhaps not exhaustive. One may try to find more time
dependent solutions and particularly some intersecting branes in
this background. A matrix string description can also be given to
brane solutions presented in this paper.


\vskip .2in

\noindent
{\bf Acknowledgments:} We would like to thank D. Bak, S. F. Hassan, and
A. Kumar for various useful discussions. The work of RRN was supported by
INFN. The work KLP was supported partially by PRIN 2004 - "Studi perturbativi
e non perturbativi in teorie quantistiche dei campi per le interazioni
fondamentali". The work of SS was supported by Korea Science Foundation
ABRL program (R14-2003-012-01001-0).



\begin{thebibliography}{99}

\bibitem{Craps:2005wd}
  B.~Craps, S.~Sethi and E.~P.~Verlinde,
  JHEP {\bf 0510} (2005) 005
  [arXiv:hep-th/0506180].


\bibitem{Banks:1996my}
  T.~Banks and N.~Seiberg,
  Nucl.\ Phys.\ B {\bf 497} (1997) 41
  [arXiv:hep-th/9702187].

\bibitem{Dijkgraaf:1997vv}
  R.~Dijkgraaf, E.~P.~Verlinde and H.~L.~Verlinde,
  Nucl.\ Phys.\ B {\bf 500} (1997) 43
  [arXiv:hep-th/9703030].

\bibitem{Bonelli:2002mb}
  G.~Bonelli,
  JHEP {\bf 0208}, 022 (2002)
  [arXiv:hep-th/0205213];
  Nucl.\ Phys.\ B {\bf 649}, 130 (2003)
  [arXiv:hep-th/0209225].


\bibitem{Li:2005sz}
  M.~Li,
  Phys.\ Lett.\ B {\bf 626}, 202 (2005)
  [arXiv:hep-th/0506260].

\bibitem{Berkooz:2005ym}
  M.~Berkooz, Z.~Komargodski, D.~Reichmann and V.~Shpitalnik,
  JHEP {\bf 0512} (2005) 018
  [arXiv:hep-th/0507067].

\bibitem{Kawai:2005jx}
  S.~Kawai, E.~Keski-Vakkuri, R.~G.~Leigh and S.~Nowling,
  Phys.\ Rev.\ Lett.\  {\bf 96}, 031301 (2006)
  [arXiv:hep-th/0507163].

\bibitem{Li:2005ti}
  M.~Li and W.~Song,
  JHEP {\bf 0510}, 073 (2005)
  [arXiv:hep-th/0507185].


\bibitem{Das:2005vd}
  S.~R.~Das and J.~Michelson,
  Phys.\ Rev.\ D {\bf 72} (2005) 086005
  [arXiv:hep-th/0508068].

\bibitem{Chen:2005mg}
  B.~Chen,
  Phys.\ Lett.\ B {\bf 632}, 393 (2006)
  [arXiv:hep-th/0508191].


\bibitem{Chen:2005bk}
  B.~Chen, Y.~l.~He and P.~Zhang,
  Nucl.\ Phys.\ B {\bf 741}, 269 (2006)
  [arXiv:hep-th/0509113].


\bibitem{Ishino:2005ru}
  T.~Ishino, H.~Kodama and N.~Ohta,
  Phys.\ Lett.\ B {\bf 631}, 68 (2005)
  [arXiv:hep-th/0509173].




\bibitem{Robbins:2005ua}
  D.~Robbins and S.~Sethi,
  JHEP {\bf 0602}, 052 (2006)
  [arXiv:hep-th/0509204].




\bibitem{Li:2005ai}
  M.~Li and W.~Song,
  arXiv:hep-th/0512335.

\bibitem{Tai:2006my}
  T.~S.~Tai,
  arXiv:hep-th/0601039.

\bibitem{Craps:2006xq}
  B.~Craps, A.~Rajaraman and S.~Sethi,
  arXiv:hep-th/0601062.



\bibitem{Chu:2006pa}
  C.~S.~Chu and P.~M.~Ho,
  arXiv:hep-th/0602054.

\bibitem{Das:2006dr}
  S.~R.~Das and J.~Michelson,
  arXiv:hep-th/0602099.

\bibitem{Das:2006dz}
  S.~R.~Das, J.~Michelson, K.~Narayan and S.~P.~Trivedi,
  arXiv:hep-th/0602107.

\bibitem{Lin:2006ie}
  F.~L.~Lin and W.~Y.~Wen,
  arXiv:hep-th/0602124.




\bibitem{Martinec:2006ak}
  E.~J.~Martinec, D.~Robbins and S.~Sethi,
  arXiv:hep-th/0603104.

\bibitem{Chen:2006rm}
  H.~Z.~Chen and B.~Chen,
  arXiv:hep-th/0603147.



\bibitem{Ishino:2006nx}
  T.~Ishino and N.~Ohta,
  arXiv:hep-th/0603215.


\bibitem{Kodama:2006bw}
  H.~Kodama and N.~Ohta,
  arXiv:hep-th/0605179.




\bibitem{Hikida:2005ec}
  Y.~Hikida, R.~R.~Nayak and K.~L.~Panigrahi,
  JHEP {\bf 0509}, 023 (2005)
  [arXiv:hep-th/0508003].

\bibitem{KalyanaRama:2005uw}
  S.~Kalyana Rama,
  arXiv:hep-th/0510008.

\bibitem{She:2005mt}
  J.~H.~She,
  JHEP {\bf 0601}, 002 (2006)
  [arXiv:hep-th/0509067];
  arXiv:hep-th/0512299.

\bibitem{Craps:2006yb}
  B.~Craps,
  arXiv:hep-th/0605199.


\bibitem{Papadopoulos:2002bg}
  G.~Papadopoulos, J.~G.~Russo and A.~A.~Tseytlin,
  Class.\ Quant.\ Grav.\  {\bf 20} (2003) 969
  [arXiv:hep-th/0211289].

\bibitem{Blau:2003rt}
  M.~Blau, M.~O'Loughlin, G.~Papadopoulos and A.~A.~Tseytlin,
  Nucl.\ Phys.\ B {\bf 673} (2003) 57
  [arXiv:hep-th/0304198].

\bibitem{Nayak:2006dm}
  R.~R.~Nayak and K.~L.~Panigrahi,
  arXiv:hep-th/0604172.

\bibitem{Schwarz:1983qr}
  J.~H.~Schwarz,
  Nucl.\ Phys.\ B {\bf 226}, 269 (1983).

\bibitem{Hassan:1999bv}
  S.~F.~Hassan,
  Nucl.\ Phys.\ B {\bf 568}, 145 (2000)
  [arXiv:hep-th/9907152].

\end{thebibliography}
\end{document}